\documentclass[showpacs,preprintnumbers,amsmath,amssymb, twocolumn,prb,superscriptaddress]{revtex4}
\usepackage[dvips]{graphics,color}
\usepackage{dcolumn}
\begin{document}
\title{Theory of Weak Localization in Ferromagnetic (Ga,Mn)As}
\author{Ion Garate}
\affiliation{Department of Physics, The University of Texas at Austin, Austin TX 78712-0264}
\author{Jairo Sinova}
\affiliation{Department of Physics, Texas A \& M University, College Station, TX 77843-4242 }
\affiliation{Institute of Physics ASCR, v.v.i., Cukrovarnicka 10, 162 53 Praha 6, Czech Republic}
\author{T. Jungwirth}
\affiliation{Institute of Physics ASCR, v.v.i., Cukrovarnicka 10, 162 53 Praha 6, Czech Republic}
\affiliation{School of Physics and Astronomy, University of Nottingham, Nottingham NG7 2RD, United Kingdom. }
\author{A.H. MacDonald}
\affiliation{Department of Physics, The University of Texas at Austin, Austin TX 78712-0264}

\date{\today}
\begin{abstract}
We study quantum interference corrections to the conductivity in (Ga,Mn)As ferromagnetic semiconductors 
using a model with disordered valence band holes coupled to localized Mn moments through a $p-d$ kinetic-exchange interaction.
We find that at Mn concentrations above 1\%
quantum interference corrections lead to negative magnetoresistance, {\em i.e.} to weak localization (WL) rather than weak antilocalization (WAL).
Our work highlights key qualitative differences between (Ga,Mn)As and previously studied 
toy model systems,
and pinpoints the mechanism by which exchange splitting in the ferromagnetic state converts valence band WAL into 
WL.  We comment on recent experimental studies and theoretical analyses of low-temperature magnetoresistance 
in (Ga,Mn)As which have been variously interpreted as implying both WL and WAL and as requiring an 
impurity-band interpretation of transport in metallic (Ga,Mn)As.
\end{abstract}

\maketitle
\section{Introduction}
At low temperatures the conductivity of disordered metals and semiconductors departs from the classical Drude formula
because of both electron-electron interaction and quantum interference corrections.\cite{WL review}
The quantum interference correction to semiclassical transport theory 
is dominated by contributions from self-intersecting paths.  The interference 
may be constructive or destructive depending on a combination of intrinsic and extrinsic factors. 
The principal extrinsic factors which help determine the sign of the quantum correction are the strength of spin-orbit (SO) scattering off heavy impurities\cite{bergmann, geller} 
and the presence (or absence) of spin-dependent scatterers. 
When SO scattering is weak and spin-dependent scatterers are absent  
interference between time reversed electron waves is constructive, which decreases conductivity and leads to weak localization (WL). 
For sufficiently strong SO scattering interference in paramagnetic systems becomes 
destructive and the conductivity is enhanced, leading to weak antilocalization (WAL). 
The main intrinsic factor which helps determine the sign of the quantum correction is helicity in 
the band structure, which is often\cite{knap,pedersen} but not always\cite{mcCan, ando} associated with spin-orbit interactions.
Heuristically, WL (WAL) is favored when the helicity of the band eigenstates is such that quasiparticles spinors at opposite momenta are 
parallel (anti-parallel).\cite{ando}
 
Both WL and WAL are suppressed by an applied magnetic field which washes out quantum interference of the carriers by effectively reducing their phase coherence length  $l_{\phi}$. 
The suppression, which is complete when the magnetic length is smaller than the quasiparticle mean free path, is the most 
common experimental signature
of the phenomenon and manifests itself as a negative (in case of WL) or positive (in case of WAL) magnetoresistance (MR).

Both the theory and the observation of WL or WAL are more complex in ferromagnetic than in paramagnetic conductors.
Experimentally, internal magnetic fields, anisotropic magnetoresistance (AMR), and isotropic magnetoresistance (IMR) can either  
destroy quantum interference or mask its occurrence.
Difficulties in interpretation can be especially severe in carrier mediated ferromagnets because of the sensitivity of quasiparticle properties to 
the magnetic microstructure. Theoretically, the role of exchange splitting when combined with intrinsic and extrinsic spin-orbit interactions
alters quantum interference in a way which has previously been incompletely articulated.
At any rate, it is agreed that neither WL nor WAL survive in clean strong ferromagnets for which the magnetic length
$l_{H}=(\hbar/(e H_{int}))^{1/2}$ is smaller than the quasiparticle mean free path $l$. (Here $H_{int}$ is the internal field of the ferromagnet.) 
On the other hand, traces of quantum interference are expected to survive when $l_{H}$ is larger than $l$.   Larger values of 
$l_{H}/l$ can be due either to weaker internal fields 
or to stronger disorder.

Because their moments are dilute and randomly distributed,
diluted magnetic semiconductors like (Ga,Mn)As have short mean-free-paths ($l \lesssim 5$ nm)  and weak internal fields ($l_{H} \gtrsim 100$ nm 
in the absence of external fields at $5$\% Mn). 
Dilute moments also help support the 
sizeable coherence lengths ($\simeq 100$ nm at $10$ mK) observed in these materials.\cite{wagner06,neumaier}
Indeed, the presence of quantum interference effects in (Ga,Mn)As has been clearly demonstrated by 
measurements of universal conductance fluctuations and Aharonov-Bohm effects in (Ga,Mn)As nanodevices.\cite{wagner06,Vila:2007_a}
However, due to the aforementioned experimental subtleties conflicting conclusions 
have been reached\cite{matsukura,neumaier, purdue} on the magnitude and even on the 
sign of quantum corrections to the conductivity.

In this paper we report on a theoretical study which we expect to be helpful in achieving a more complete understanding.
Unlike earlier theoretical work\cite{dugaev, sil} which addressed quantum interference in ferromagnets, we focus our study 
on a four-band model which is directly relevant to the valence bands of (Ga,Mn)As.
We demonstrate that the quantum interference contribution to 
MR in robustly ferromagnetic (Ga,Mn)As is negative.  Our theoretical conclusion 
is at odds with the outcome of the experimental study of Neumaier {\em et al.},\cite{neumaier}
and in partial agreement with the purported conclusion of the experimental study of Rokhinson {\em et al.}.\cite{purdue}  
As we discuss later, however, it is not yet completely clear that either experimental study has completely succeeded 
in separating the quantum interference correction to the semiclassical conductivity from other 
magnetoresistance effects, the most troubling of which is likely anisotropic magnetoresistance.
In their experimental study of magnetoresistance in (Ga,Mn)As, Rokhinson {\em et al.} argued  
that their observation of negative MR is incompatible with quantum 
interference theory, and that it therefore implied that transport 
must be occurring within an impurity band.  
The opposite conclusion, namely that ferromagnets should normally exhibit WL rather than WAL was reached in an earlier theoretical contribution by Dugaev {\em et al.}.\cite{dugaev} Nevertheless, in a recent comment\cite{brunocomment} Dugaev {\em et al.} have explained that their theory can also lead to WAL under some circumstances, arguing that it is not necessarily at odds with the Neumaier {\em et al.} WAL finding in (Ga,Mn)As.
As the detailed theory presented in this paper makes clear, Rokhinson {\em et al.}
are incorrect in asserting that WL cannot occur in ferromagnetic (Ga,Mn)As.

In this contribution we attempt to reduce the level of confusion by studying a model
which is directly relevant to (Ga,Mn)As and by drawing attention to some of the complications which arise in 
separating quantum interference from other MR effects.  Negative MR (WL) is   
in fact readily compatible with transport in a disordered exchange-split valence band, even when there is strong SO coupling in the band,
and {\em is expected} in (Ga,Mn)As.  Existing experimental work{\cite{neumaier,purdue} which isolates a low-temperature contribution to 
the MR of (Ga,Mn)As is strongly suggestive of a quantum interference effect.  However, additional work will be 
needed to make this identification conclusive and to test theory quantitatively by accurately isolating the quantum interfernce
contribution to MR.   
 
This paper is organized as follows. We begin by reviewing the experimental studies of low-temperature magnetoresistance phenomena in (Ga,Mn)As (Section \ref{experiments}).
We follow in Section III with an outline of the general formalism used here to evaluate the Cooperon in multi-band ferromagnets. 
In Section IV we apply the formalism to the two dimensional electron gas ferromagnet (M2DEG) model studied earlier by Dugaev {\em et al.}.\cite{dugaev}
The reults in this section are useful in discussing the competition between WL and WAL in ferromagnets generally.
Section V is devoted to the more complicated 4-band Kohn-Luttinger model with a kinetic-exchange mean-field, 
which captures the essentials\cite{RMP} of ferromagnetism in (Ga,Mn)As.
We find that in this model, which employs the disordered valence-band picture of states near the Fermi level in metallic (Ga,Mn)As and 
typically overestimates the effect of SO interactions, very small exchange fields are sufficient to convert the positve MR (WAL) of 
the paramagnetic state to negative MR (WL) in the ferromagnetic state.
WL is predicted over the entire broad range of Mn concentrations 
for which robust metallic ferromagnetism occurs in high quality (Ga,Mn)As samples with a 
low-density of Mn interstitials.
In the M2DEG model, on the other hand, relatively large exchange fields are necessary to convert WAL into WL.  
We explain in Section V that this difference in WL behavior is due to a difference in quasiparticle chirality 
between the two models.  Motivated by the large semiclassical AMR effects in (Ga,Mn)As 
which typically occur over a field range similar to that over which WL(WAL) MR effects occur, 
we explore anisotropy in the weak localization effect itself in Section VI.   
Section VII summarizes our work and highlights our principle conclusions.

\section{Review of experimental results}
\label{experiments}
As mentioned above, the measurement of WL or WAL in magnetic systems is subtle because (i) the internal magnetization of a ferromagnet partially dephases quantum interference, and 
(ii) quantum interference must be isolated from a non trivial background of semiclassical MR effects.
The background is usually primarily due to magnetization direction rotation combined with AMR 
(dependence of resistance on magnetization direction) but 
may involve field-dependent changes in magnetic microstructure ({\em e.g.} domain wall distribution)
or field-dependent spin-disorder scattering.  
Due in part to these subtleties experimental studies have reached 
contradictory conclusions about WL in (Ga,Mn)As.  In this section we briefly review and comment on the experimental literature. 
The detailed theoretical 
analysis presented in the following sections was motivated both by
experimental confusion, and by confusion about the theory 
which should be used to guide its interpretation.

Matsukura {\em et. al}\cite{matsukura} initiated the discussion of possible WL or WAL contributions to 
transport in (Ga,Mn)As by identifying it as a possible source of the isotropic negative MR which is frequently observed in (Ga,Mn)As films. 
This negative MR does not saturate even at very high magnetic fields, which may rule out the suppression of spin disorder as the responsible factor. 
At the time of these measurements there were no studies of coherence length scales in (Ga,Mn)As, and it was
then plausible that the observed phenomenon be a manifestation of weak localization.\cite{kramer:1993_a} 
In fact, the data seems to bear a reasonable fit to a $H^{1/2}$ dependence at high magnetic field $H$, which is characteristic of three-dimensional (3D) 
WL  in systems with negligible SO coupling. Given that the intrinsic SO interaction in the valence band is by no means negligible, the authors speculated
that WL occurs rather than WAL because spin-orbit coupling is rendered largely inefficient by the exchange splitting. 
However, no explicit calculation was provided to support this view.
This interpretation of the MR 
signal requires that the magnetic length scale $l_H$ be shorter than the coherence length scale $l_\phi$ at that field range. However,
Ref. \onlinecite{matsukura} includes measurements only at temperatures (above 2 K) for which $l_{\phi}$ is expected,
based on more recent experimental work,\cite{wagner06,Vila:2007_a}
to be significantly smaller than $l_H$.  The MR effect studied by Matsukura {\em et al.}\cite{matsukura} is therefore 
unlikely to be due to quantum interference.

Neumaier {\em et al.} have studied magnetotransport 
in (Ga,Mn)As by measuring the transport properties of arrays of nanowires at temperatures down to $\sim 10$~mK.\cite{neumaier} 
Like Matsukura {\em et al.} they observe a high field negative MR observed in their samples which is
ascribed either to quantum interference or to the suppression of spin disorder but is not analyzed in detail. 
Closer inspection of the Neumaier {\em et al.} data reveals that the temperature dependence of this high field MR signal 
tracks the temperature dependence of the conductivity.  Since quantum interference should be stronger in 
more disordered samples, WL is unlikely to be the origin of this MR effect. 

The main focus of Neumaier {\it et al.}'s work is on the low-field regime.  At high temperatures
a positive MR effect, due to AMR, is visible at fields below $\sim 0.4$T. 
This low-field signal in their data is clearly altered at low temperatures,  
likely because of quantum interference corrections to the conductivity.  Neumaier {\em et al.} 
ascribe the low-temperaure MR effect to WAL, but this conclusion is subject to uncertainty as we now explain.
Both low-field effects appear on top of the broader-field background mentioned above.
The temperature-dependence of the background is likely due to electron-electron interactions,
but its negative MR field-dependence is of uncertain origin.
Attempts to isolate the WAL signal by taking the difference between low-temperature and higher-temperature MR curves
are complicated by the substantial temperature dependence of the background at both low and high fields.
Neumaier {\it et al.} choose to identify the quantum interference contribution by examining the 
temperature dependence of the ratio of low-field and high-field resistances, and in this way conclude that 
it has WAL character, {\em i.e.} that the correction gives a positive contribution to the low field magnetoresistance 
However, this interpretation is somewhat fragile because changes in internal magnetization at these low fields
can yield resistance changes which are difficult to accurately anticipate.\cite{footnote}
Other contributions to MR can play an important role, can be temperature dependent, and are
therefore not uniquely separable from the WL/WAL effects.
As noted by Neumaier {\it et al.} , the experimentally extracted contribution ascribed to the WAL saturates at lower magnetic fields than expected from the inferred $l_{\phi}$ and $l$. 
These saturation fields are similar to the magnetic anisotropy fields in the material, suggesting that 
the present interpretation is not complete.\cite{dimensions} 
Our theoretical results in Sections V and VI suggest rather strongly that WAL is in fact not expected to prevail in metallic (Ga,Mn)As.  

Although our calculations do not support Neumaier {\em et al.}'s\cite{neumaier} WAL interpretation 
of the positive low-field MR, they do not provide an immediate alternative interpretation of the data.
We explore one possibility in Section VI by theoretically examining the anisotropy of the 
quantum interference corrections.  In our theory the symmetry breaking mechanism for this anisotropy is the same
as that responsible for the higher-temperature AMR effect.  We conclude that anisotropy of the WL corrections to conductivity also cannot account for the changes in MR observed in experiment at low temperatures. It is possible that still more elaborate experimental studies with magnetic fields applied along different directions, and using materials with both 
in-plane and out-of-plane magnetic easy-axes and various nano-bar geometries will be able to separate AMR and quantum interference effects to 
achieve a complete picture of low-temperature MR in (Ga,Mn)As. Effects that may compete with the WL/WAL  include higher order (cubic, etc.) AMR terms, 
spin-disorder scattering, or electron-electron interactions.

Finally we would like to comment on the work by Rokhinson {\em et al.} in which they observe
an intriguing MR
peak at low temperature and low fields in (Ga,Mn)As films\cite{purdue}
which is interpreted as a signature of the WL. Although our results in Sections V and VI may appear to corroborate this experimental work, the character and field-range of the measured negative MR 
seems to be incompatible with WL theory.   
In particular,  the expected higher-field $\sim H^{1/2}$ MR tail is not seen in the measured data. 
Also the experimental dependence of the  MR on the orientation of the magnetic field  is not understood theoretically.
Further experiments and analyses will be necessary to conclusively establish the origin of the low-temperature
MR seen in these experiments.  

Our theoretical work does 
directly address the qualitative conclusion drawn in Ref. ~[\onlinecite{purdue}] concerning the nature of Fermi level states in metallic (Ga,Mn)As. 
Rokhinson {\em et al.} argue on qualitative grounds that the WL quantum interference effect they apparently observe 
is not compatible with conduction in a disordered valence band.
They then leap to the conclusion that the itinerant holes in (Ga,Mn)As must be in an impurity band, arguing that spin-orbit interactions would be weaker in that case.
In doing so they are connecting with an issue which has been controversial\cite{kenny,sbreview,tomasib,ericyib} 
in the (Ga,Mn)As literature, namely whether transport electrons in metallic (Ga,Mn)As 
should be viewed as being in a valence band or an impurity band. 
Although there is a sharp distinction between metallic and 
insulating behavior, there is in fact no sharp distinction between a disordered valence band and 
an impurity band in a semiconductor.  In the present 
context the statement that the transport electrons are in an impurity band presumably is a statement 
that the scattering potential between valence band electrons and the Mn impurities is 
sufficiently strong that the relevant Hilbert space can be obtained by projecting the direct product of 
isolated impurity acceptor levels from the valence band.  Presumably the perturbative treatment 
of disorder used in quantum interference theory would then be invalid.
Although there is no theory for quantum interference in the impurity band limit, 
Rokhinson {\em et al.} nevertheless argue that impurity-band conduction  
might explain their observation of WL instead of the WAL they expect.
Since, as we show in Sections V and VI, standard quantum interference theory in the disordered
SO-coupled exchange-split valence band of (Ga,Mn)As implies WL not WAL, the experimental
finding of Rokhinson {\em et al.} is in fact perfectly consistent with transport in a valence
band with disorder which can be treated perturbatively.   

\begin{widetext}
\section{Evaluation of the Cooperon in Conducting Ferromagnets}
In this section we present a  formalism to evaluate quantum interference corrections to conductivity in multi-band, disordered ferromagnets with intrinsic SO interactions. 
In the diffusive regime ($l_{H}, l_{\phi} \gg l$) these corrections are captured by a geometric sum of the maximally crossed diagrams,\cite{WL review} which are encoded in the so-called Cooperon $C$ (Fig. 1).
\begin{figure}[h]
\begin{center}
\scalebox{0.5}{\includegraphics{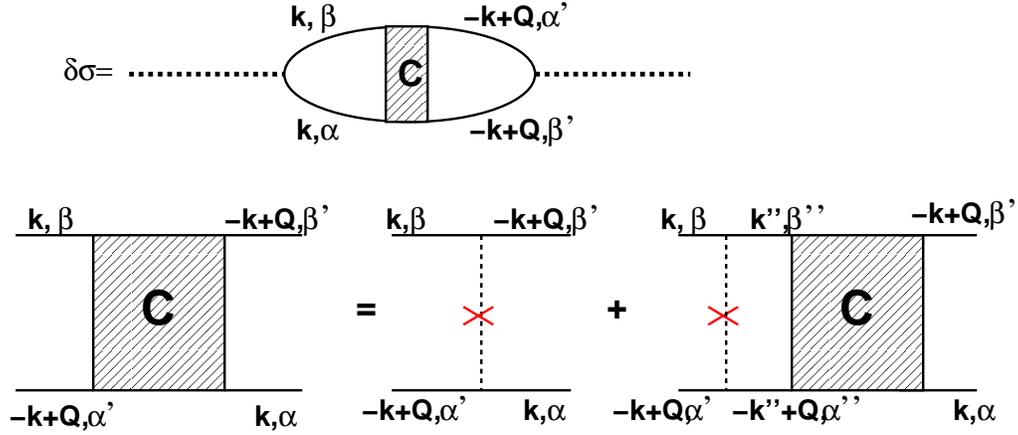}}
\caption{Diagrammatic expressions for the quantum interference 
correction to the conductivity and the Cooperon. Crosses stand for scattering events off impurities.}
\label{fig:figure}
\end{center}
\end{figure}
The deviations from the Drude conductivity may be read out from Fig. 1, i.e.
\begin{equation}
\label{eq:delta_sigma}
\delta\sigma=\frac{e^{2}}{2\pi}\int d{\textbf{k} } \; v_{\alpha,\beta}^{x}(\textbf{k})v_{\beta^{\prime},\alpha^{\prime}}^{x}(-\textbf{k)}G_{\alpha}^{R}(\textbf{k})G_{\alpha^{\prime}}^{R}(-\textbf{k})G_{\beta}^{A}(\textbf{k})G_{\beta^{\prime}}^{A}(-\textbf{k})\int d{\textbf{Q}} \;  C^{\beta,\beta^{\prime}}_{\alpha^{\prime},\alpha}(\textbf{k},\textbf{Q})
\end{equation}
\end{widetext}
where we set $\hbar\equiv 1$, $\alpha,\beta, ...$ label band eigenstates of the ferromagnet, $\textbf{v}$ is the carrier velocity operator, $v_{\alpha\beta}^{i}(\textbf{k}_{1},\textbf{k}_{2})\equiv\langle\alpha\textbf{k}_{1}|v^{i}|\beta\textbf{k}_{2}\rangle$, $G^{A(R)}$ is the advanced (retarded) Green's function in the first Born approximation, and $\textbf{Q}$ is the ``center of mass'' momentum of the Cooperon. 
$Q$ ranges approximately from the inverse phase coherence length, $1/l_{\phi}$, to the inverse mean free path, $1/l$. 
Following standard practice, we have kept \textbf{Q} in the Cooperon propagator only in Eq.(~\ref{eq:delta_sigma}), setting $Q=0$ elsewhere in the integrand.
Additionally, we ignore the contribution from non-backscattering processes,\cite{dmitriev, golub} which are unimportant in the diffusive regime.

The main challenge resides in evaluating $C$, which obeys the Bethe-Salpeter equation (see Fig.1):
\begin{widetext}
\begin{eqnarray}
\label{eq:Dyson}
C^{\beta,\beta^{\prime}}_{\alpha^{\prime},\alpha}(\textbf{k},\textbf{Q})&=& u^{a} J^{a}_{\beta,\beta^{\prime}}(\textbf{k},-\textbf{k}+\textbf{Q})J^{a}_{\alpha^{\prime},\alpha}(-\textbf{k}+\textbf{Q},\textbf{k})\nonumber\\
&+&\int d{\textbf{k}^{\prime\prime}} u^{a}J^{a}_{\beta,\beta^{\prime\prime}}(\textbf{k},\textbf{k}^{\prime\prime}) J^{a}_{\alpha^{\prime},\alpha^{\prime\prime}}(-\textbf{k}+\textbf{Q},-\textbf{k}^{\prime\prime}+\textbf{Q}) G^{A}_{\beta^{\prime\prime}}(\textbf{k}^{\prime\prime})G^{R}_{\alpha^{\prime\prime}}(-\textbf{k}^{\prime\prime}+\textbf{Q}) C^{\beta^{\prime\prime},\beta^{\prime}}_{\alpha^{\prime\prime},\alpha}(\textbf{k},\textbf{k}^{\prime\prime},\textbf{Q}).
\end{eqnarray}
\end{widetext}
In Eq.(~\ref{eq:Dyson}) we include short-range disorder potentials which can differ between the ferromagnet's 
majority and minority spins but do not include spin-flip disorder which would not play a distinct role 
because we retain spin-orbit coupling in the band structure.  Accordingly  
$\textbf{J}$ is the carrier spin-density operator (with component $a=0$ reserved for charge-density) and 
$J_{\alpha\beta}^{i}(\textbf{k}_{1},\textbf{k}_{2})\equiv\langle\alpha\textbf{k}_{1}|J^{i}|\beta\textbf{k}_{2}\rangle$ 
is the matrix element of this operator between Bloch states.  In our calculations we include only a 
$a=0$ spin-independent contribution to the disorder potential which we refer to as Coulomb scattering 
and a $a=z$ spin-dependent contribution which we refer to as magnetic impurity scattering.
($\hat{z}$ is the direction of magnetization.) 
A sum over $a,\alpha^{\prime\prime},\beta^{\prime\prime}$ is implicit.
The quasiparticle lifetimes are given by Fermi's golden rule:
\begin{equation}
\frac{1}{\tau_{{\bf k},\alpha}}=2\pi\int d{\textbf{k}^{\prime}}u^{a}\sum_{\alpha^{\prime}}J^{a
}_{\alpha,\alpha^{\prime}} J^{a}_{\alpha^{\prime},\alpha} \delta(E_{\textbf{k}\alpha}-E_{\textbf{k}^{\prime}\alpha^{\prime}})
\end{equation}
where $u^{a}\equiv n_{a} V_{a}^{2}({\bf q=0})$ for $a\in\{0,z\}$, $n_{a}$ is the density of scatterers and $V_{a}(\textbf{q})$ is the scattering potential (dimensions: $\mbox{(energy)}\times\mbox{(volume)}$). This model of disorder assumes independent incoherent scattering of Coulomb and magnetic impurities and is sufficient for the main purposes of this work. However, in Section VI we shall consider the case of coherent magnetic and non-magnetic scattering, which more faithfully describes the nature of the randomly distributed substitutional Mn impurities in (Ga,Mn)As 
which carry local moments exchange coupled to the holes and are charged acceptors.  Accounting for correlations between Coulomb and spin-dependent scattering  
allows us to asses the strength of the WL and WAL contributions, including their anisotropies with respect to the magnetization orientation,\cite{purdue} on a more quantitative level.

Eq. (\ref{eq:Dyson}) is an integral equation of considerable complexity, mainly due to the $\textbf{k}^{\prime\prime}$-dependence of the Cooperon inside the integrand. Rather than attempting to solve it fully numerically, we shall proceed to simplify Eq. (~\ref{eq:Dyson}) analytically. The simplest approach would be to try an ansatz in which the Cooperon depends only on the center of mass momentum rather than on the incoming and outgoing momenta separately; however, this ansatz fails whenever the eigenstates of the ferromagnet are momentum-dependent. 
The approach we use is based fundamentally on the property that the disorder potentials is local.  On each site the 
disorder potential can be expressed in any representation for the bands included in the model, for example the four bands 
included in the $J=3/2$ Kohn-Luttinger valence band model. 
We therefore express matrix element of the spin operator in the basis spanned by $J^{z}$ eigenstates:\cite{other_bases}
\begin{widetext}
\begin{equation}
\label{eq:Jz_expansion}
J^{a}_{\alpha,\alpha^{\prime}}=\langle\alpha,\textbf{k}|J^{a}|\alpha^{\prime},-\textbf{k}+\textbf{Q}\rangle=\sum_{m,m^{\prime}}\langle\alpha,
\textbf{k}|m\rangle\langle m^{\prime}|\alpha^{\prime},-\textbf{k}+\textbf{Q}\rangle \langle m|J^{a}|m^{\prime}\rangle
\end{equation}
where $|m\rangle$ satisfies $J^{z}|m\rangle=m|m\rangle$. Now the entire $\textbf{k}$-dependence of $J^{a}_{\alpha,\alpha^{\prime}}$ is contained on $\langle\alpha,\textbf{k}|m\rangle\langle m^{\prime}|\alpha^{\prime},-\textbf{k}+\textbf{Q}\rangle$. In the same spirit, we can decompose $C$ in the $J^{z}$ basis:
\begin{eqnarray}
\label{eq:expanded_Cooperon}
C^{\beta,\beta^{\prime}}_{\alpha^{\prime},\alpha}(\textbf{k},\textbf{Q})&=&\left(\langle\alpha,-\textbf{k}+\textbf{Q}|\otimes\langle\beta,\textbf{k}|\right)C\left(|
\beta^{\prime},-\textbf{k}+\textbf{Q}\rangle\otimes|\alpha,\textbf{k}\rangle\right)\nonumber\\
&=& \sum_{m,m^{\prime},n,n^{\prime}}\langle\alpha^{\prime},-\textbf{k}+\textbf{Q}|m^{\prime}\rangle\langle\beta,\textbf{k}|m\rangle\langle n|
\beta^{\prime},-\textbf{k}+\textbf{Q}\rangle\langle n^{\prime}|\alpha\textbf{k}\rangle C^{m,n}_{m^{\prime},n^{\prime}}(\textbf{Q}).
\end{eqnarray}
The critical property which simplifies our calculation is the observation that 
$ C^{m,n}_{m^{\prime},n^{\prime}}$ depends \emph{only} on $\textbf{Q}$ in the case of local disorder; 
the dependence on $\textbf{k}$ is captured by the overlap matrix elements in Eq. (~\ref{eq:expanded_Cooperon}). Applying the same transformation to the terms on the right hand side of Eq. (~\ref{eq:Dyson}) and eliminating common factors on both sides we arrive at
\begin{equation}
\label{eq:iterative}
C^{m,n}_{m^{\prime},n^{\prime}}(\textbf{Q})=(u^{0}+m m^{\prime} u^{z})\delta_{m,n}\delta_{m^{\prime},n^{\prime}}+\sum_{l,l^{\prime}}U^{m,l}_{m^{\prime},l^{\prime}}(\textbf{Q}) C^{l,n}_{l^{\prime},n^{\prime}}(\textbf{Q})
\end{equation}
where
\begin{equation}
\label{eq:U}
U^{m,l}_{m^{\prime},l^{\prime}}(\textbf{Q})=(u^{0}+m m^{\prime} u^{z})\int d{\textbf{k}}\langle m|\alpha\textbf{k}\rangle G^{A}_{\alpha}(\textbf{k})\langle\alpha\textbf{k}|l\rangle
\langle m^{\prime}|\alpha^{\prime},-\textbf{k}+\textbf{Q}\rangle G^{R}_{\alpha^{\prime}}(-\textbf{k}+\textbf{Q})\langle\alpha^{\prime},-\textbf{k}+\textbf{Q}|l^{\prime}\rangle.
\end{equation}
\end{widetext}
Eq. (~\ref{eq:iterative}) may be conveniently expressed in matrix form
\begin{equation}
\label{eq:matrix equation}
\textbf{C}=(\textbf{1}-\textbf{U})^{-1}\textbf{C}^{(0)},
\end{equation}
where $C^{(0)}=(u^{0}+m m^{\prime} u^{z})\delta_{m,n}\delta_{m^{\prime},n^{\prime}}$ in the $J^{z}$ representation.
From Eq. (~\ref{eq:matrix equation}), we see that the  
quantum interference correction to conductivity is largely governed by modes or channels for 
which the eigenvalues of $U({\bf Q})$ are equal to (or closest to) one.  
For the models we study, and in almost any physically realistic situation, 
the largest eigenvalues of $\textbf{U}$ are smaller than or equal to $1$ and occur at $Q=0$.
The $Q$-dependence of all eigenvalues is given approximately by $ - D Q^2$ where $D$ is the diffusion constant.
It follows that the main contribution to quantum interference comes from small \textbf{Q} region. 
Thus it is appropriate to simplify Eq. (~\ref{eq:U}) as
\begin{widetext}
\begin{equation}
\label{eq:U_expanded}
U^{m,l}_{m^{\prime},l^{\prime}}\simeq(u^{0}+m m^{\prime} u^{z})\int_{\textbf{k}}\langle m|\alpha\textbf{k}\rangle G^{A}_{\alpha}(\textbf{k})\langle\alpha\textbf{k}|l\rangle
\langle m^{\prime}|\alpha^{\prime},-\textbf{k}\rangle G^{R}_{\alpha^{\prime}}(\textbf{k})\langle\alpha^{\prime},-\textbf{k}|l^{\prime}\rangle\left[1+(\textbf{v}_{\textbf{k},\alpha^{\prime}}\cdot\textbf{Q}) G^{R}_{\alpha^{\prime}}(\textbf{k})+ (\textbf{v}_{\textbf{k},\alpha^{\prime}}\cdot\textbf{Q})^{2}G^{R}_{\alpha^{\prime}}(\textbf{k})^{2}\right]
\end{equation}
\end{widetext}
where $\textbf{v}_{\textbf{k},\alpha}=\partial{E_{\textbf{k},\alpha}}/\partial{\textbf{k}}$ is the quasiparticle velocity. In Eq. (~\ref{eq:U_expanded}) we have assumed that the momentum dependence of the scattering rate is small, have omitted O($Q^{2}$) terms that are negligible for $k_{F} l \gg 1$, and have made use of $G(-\textbf{k})=G(\textbf{k})$. Moreover, we have neglected the \textbf{Q}-dependence of the eigenstates, because the most singular behavior originates from the \textbf{Q}-dependence of the Green's functions.  
The combination of Eqs. (~\ref{eq:delta_sigma}), (~\ref{eq:expanded_Cooperon}), (~\ref{eq:matrix equation}) and (~\ref{eq:U_expanded}) yields the complete solution to the problem in hand. In the 
following sections we apply them to the two-band M2DEG toy-model ferromagnet studied previously be Dugaev {\em et al.}\cite{dugaev} and to the 4-band, Kohn-Luttinger kinetic-exchange model of 
ferromagnetic (Ga,Mn)As.

\section{Quantum Corrections to Conductivity in a Magnetized 2DEG}

A magnetized two-dimensional electron gas (M2DEG) is a minimal model to describe a ferromagnet with intrinsic SO interactions. 
Altough it is overly simplistic, the M2DEG model provides a versatile theoretical platform where the formalism developed in the previous section may be tested while at the same time gaining insight on how exchange fields, intrinsic SO and quasiparticle chirality influence quantum interference.\cite{dugaev} From a more pragmatic standpoint, a M2DEG is known to capture the qualitative features of magnetization relaxation in ferromagnetic semiconductors both in presence and absence of a transport current;\cite{our_papers} it is then not unreasonable to expect a similar qualitative guidance regarding weak localization.\cite{wrong_hope} 

The Hamiltonian that describes an M2DEG is
\begin{equation}
H=\frac{k^{2}}{2 m}+\textbf{b}_{\textbf{k}}\cdot\textbf{J}
\end{equation}
where $\textbf{b}_{\textbf{k}}=(-\lambda k_{y}, \lambda k_{x}, h_{z})$, $h_{z}$ is the exchange field (perpendicular to the 2DEG), $\lambda$ is the Rashba SO coupling and ${\bf J}$ is the spin operator for spin 1/2 quasiparticles. The corresponding eigenvalues and eigenstates are
\begin{eqnarray}
\label{eq:eigen_M2DEG}
&& E_{\pm,\textbf{k}}=\frac{k^2}{2m}\pm \sqrt{h_{z}^{2}+\lambda^{2}k^{2}}\nonumber\\
&&|\alpha\textbf{k}\rangle = e^{-i J^{z}\phi} e^{-i J^{y}\theta}|\alpha\rangle
\end{eqnarray}
where $\phi=-\tan^{-1}(k_{x}/k_{y})$ and $ \theta=\cos^{-1}(h_{z}/\sqrt{h_{z}^{2}+\lambda^{2}k^{2}})$ are the spinor angles and $\alpha=\pm$ is the band index.
Given Eq.(~\ref{eq:eigen_M2DEG}), one may attempt to solve Eqs. (~\ref{eq:delta_sigma}), (~\ref{eq:expanded_Cooperon}), (~\ref{eq:matrix equation}) and (~\ref{eq:U_expanded}). Due to the rotational symmetry in the plane of the 2DEG, the azimuthal integration in Eq.(~\ref{eq:U_expanded}) yields $U^{m,l}_{m^{\prime},n^{\prime}}(0)\propto \delta_{m+m^{\prime},l+l^{\prime}}$, whereupon we obtain
\begin{widetext}
\begin{equation}
\label{eq:U_M2DEG}
\textbf{U(0)}=\left(\begin{array}{cccc} U^{\uparrow\uparrow}_{\uparrow\uparrow} & 0 & 0 & 0\\
                          0 & U^{\uparrow\uparrow}_{\downarrow\downarrow} & U^{\uparrow\downarrow}_{\downarrow\uparrow} & 0\\
                          0 & U^{\downarrow\uparrow}_{\uparrow\downarrow} & U^{\downarrow\downarrow}_{\uparrow\uparrow} & 0\\
                          0 & 0 & 0 & U^{\downarrow\downarrow}_{\downarrow\downarrow}
\end{array}\right)
\end{equation}   
in the $\{|m_{1},m_{2}\rangle\}=\{|\uparrow\uparrow\rangle, |\uparrow\downarrow\rangle, |\downarrow\uparrow\rangle, |\downarrow\downarrow\rangle\}$ basis. Assuming that the band splitting is small ($\lambda k_{F},h_{z}<<\epsilon_{F}$), the matrix elements of Eq.(~\ref{eq:U_M2DEG}) may be obtained analytically:
\begin{eqnarray}
\label{eq:U_analytical}
U^{\uparrow\uparrow}_{\uparrow\uparrow}(0)&=&\cos^{4}\left(\frac{\theta}{2}\right)+\sin^{4}\left(\frac{\theta}{2}\right)
+\frac{i\Gamma}{-2 b+i\Gamma}\cos^{2}\left(\frac{\theta}{2}\right)\sin^{2}\left(\frac{\theta}{2}\right)
+\frac{i\Gamma}{2 b+i\Gamma}\cos^{2}\left(\frac{\theta}{2}\right)\sin^{2}\left(\frac{\theta}{2}\right)\nonumber\\
U^{\downarrow\downarrow}_{\downarrow\downarrow}(0)&=&U^{\uparrow\uparrow}_{\uparrow\uparrow}(0)\nonumber\\
U^{\uparrow\uparrow}_{\downarrow\downarrow}(0)&=&\frac{u^{0}-u^{z}/4}{u^{0}+u^{z}/4}\left[2\cos^{2}\left(\frac{\theta}{2}\right)\sin^{2}\left(\frac{\theta}{2}\right)
+\frac{i\Gamma}{-2b+i\Gamma}\cos^{4}\left(\frac{\theta}{2}\right)
+\frac{i\Gamma}{2b+i\Gamma}\sin^{4}\left(\frac{\theta}{2}\right)\right]\nonumber\\
U^{\downarrow\downarrow}_{\uparrow\uparrow}(0)&=&\left[U^{\uparrow\uparrow}_{\downarrow\downarrow}(0)\right]^{\star}\nonumber\\
U^{\uparrow\downarrow}_{\downarrow\uparrow}(0)&=&\frac{u^{0}-u^{z}/4}{u^{0}+u^{z}/4}\left(-2
+\frac{i\Gamma}{-2b+i\Gamma}
+\frac{i\Gamma}{2b+i\Gamma}\right)\cos^{2}\left(\frac{\theta}{2}\right)\sin^{2}\left(\frac{\theta}{2}\right)\nonumber\\
U^{\downarrow\uparrow}_{\uparrow\downarrow}(0)&=&\left[U^{\uparrow\downarrow}_{\downarrow\uparrow}(0)\right]^{\star}=U^{\uparrow\downarrow}_{\downarrow\uparrow}(0)
\end{eqnarray}
\end{widetext}
where $b=\sqrt{h_{z}^{2}+\lambda^{2} k_{F}^{2}}$ and $\Gamma=1/\tau$ are the band splitting and the scattering rate at the Fermi energy. We are ultimately interested in the $Q\neq 0$ generalization of Eq.(~\ref{eq:U_analytical}), which is straightforward if we assume $U^{m,l}_{m^{\prime},l^{\prime}}(Q)\propto \delta_{m+m^{\prime},l+l^{\prime}}$. 
While this assumption neglects the departure from azimuthal symmetry at $Q\neq 0$, we find it to be in good quantitative agreement with the full numerical calculation.
With this proviso, integration of Eq.(~\ref{eq:U_expanded}) yields
\begin{equation}
\label{eq:U_Q_M2DEG}
U^{m,l}_{m^{\prime},l^{\prime}}(Q)\simeq U^{m,l}_{m^{\prime},l^{\prime}}(0)(1-D Q^{2}\tau)
\end{equation}
where $D=v_{F}^{2}\tau/2$ is the diffusion constant in 2D. Substituting Eqs.(~\ref{eq:U_analytical}) and (~\ref{eq:U_Q_M2DEG}) in Eq.(~\ref{eq:matrix equation}) it is feasible to derive concise analytical expressions for  Eq.(~\ref{eq:delta_sigma}) in limiting cases; the following discussion and Table 1 summarizes the results.  
These calculations are described in greater detail in Appendix A. 

\begin{table}[t]
\caption{Quantum correction to conductivity in simple limits of the M2DEG model. 
We have set the the external magnetic field and $u^{z}$ to zero in constructing this table.
 $\delta\sigma_{0}\equiv \frac{e^{2}}{2\pi^{2}}\log\frac{\tau_{\phi}}{\tau}$ and $\Gamma = 1/\tau$ is the 
band eigenstate energy uncertainty, which is assumed to be smaller than the Fermi energy.}
\vspace{0.1 in}
\begin{tabular}{c c c}
\hline\hline
    & $\delta\sigma/\delta\sigma_{0}$ & singular channel(s) \\
\hline\\
$\lambda=h_{z}=0$ & -1 & $|\uparrow\uparrow\rangle,|\downarrow\downarrow\rangle,|\uparrow\downarrow\rangle,|\downarrow\uparrow\rangle$
\\[1ex]
\hline\\
$0<h_{z}<\Gamma$\\ $\lambda=0$ & -1 & $|\uparrow\uparrow\rangle,|\downarrow\downarrow\rangle,|\uparrow\downarrow\rangle,|\downarrow\uparrow\rangle$
\\[1ex]
\hline\\
$0<\lambda k_{F}<\Gamma$\\ $h_{z}=0$ & -1 & $|\uparrow\uparrow\rangle,|\downarrow\downarrow\rangle,|\uparrow\downarrow\rangle,|\downarrow\uparrow\rangle$
\\[1ex]
\hline\\
$\Gamma<h_{z}<<\epsilon_{F}$\\ $\lambda=0$ & -1 & $|\uparrow\uparrow\rangle,|\downarrow\downarrow\rangle$
\\[1ex]
\hline\\
$\Gamma<\lambda k_{F}<<\epsilon_{F}$\\$ h_{z}=0$ & $\frac{1}{2}$ & $\frac{1}{\sqrt{2}}\left(|\uparrow\downarrow\rangle-|\downarrow\uparrow\rangle\right)$
\\[1ex]
\hline
\end{tabular}
\end{table}
\begin{figure}[h]
\begin{center}
\scalebox{0.45}{\includegraphics{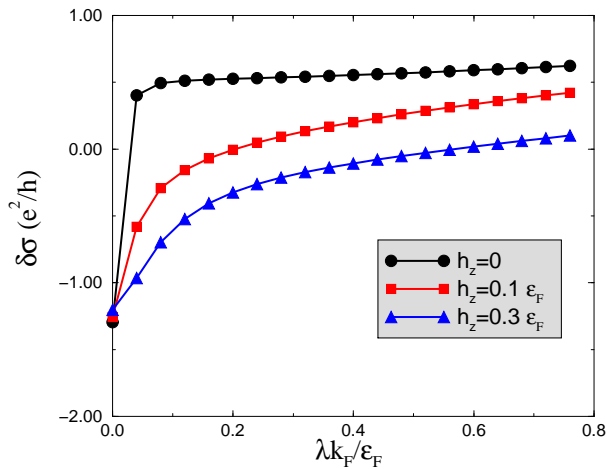}}
\caption{(color online) \textbf{M2DEG}: Quantum correction to conductivity in the absence of an external magnetic field
for $\epsilon_{F}\tau=20$, $\tau_{\phi}\simeq 42 \tau$, and $u^{z}=0$. There is a WAL-to-WL crossover when
the exchange field becomes comparable to the SO interaction strength. (Note that the band-splitting due to the exchange field
is $2 h_{z}$).  In limiting cases there is good agreement with the expressions of Table 1. 
The fact that $\delta\sigma(\lambda=0)$ is weakly dependent on $h_{z}$ indicates that an exchange splitting \emph{per se} 
does not suppress weak localization unless it becomes comparable to $\epsilon_{F}$.
We have separately evaluated the $h_{z}=0$ curve for a different values of $l_{\phi}$ (not shown) and found a good 
agreement with Fig. 1 of Ref. [~\onlinecite{geller}].} 
\label{fig:sigma_vs_so_M2DEG}
\end{center}
\end{figure}

When $\lambda k_{F}=h_{z}=0$, \textbf{U} is diagonal in the $|m,m^{\prime}\rangle$ basis and all 4 eigenvalues are unity (for $Q=0$ and $u^{z}=0$). Ultimately only two of the modes ($|\uparrow\uparrow\rangle$ and $|\downarrow\downarrow\rangle$) contribute to the conductivity correction because, in the absence of SO, conductivity corrections originate from spin-up and spin-down carriers independently. The resulting WL expression agrees with long established results:\cite{WL review} $\delta\sigma=-\frac{e^{2}}{2\pi^{2}}\log\frac{\tau_{\phi}}{\tau}$, where $\tau_{\phi}=l_{\phi}^{2}/D$ is the coherence time.  This result remains unchanged when $ 0\neq h_{z}<<\Gamma$ and $0\neq \lambda k_F<1/(\tau \tau_{\phi})^{1/2}$, because the broadening of energy levels overcomes the band splitting and the
SO induced rotation of spins in momentum space is averaged out by momentum scattering. 

On the other hand, when $h_{z}>\Gamma$ and $\lambda k_{F}=0$, \textbf{U} is still diagonal in the $|m,m^{\prime}\rangle$ basis but only two eigenvalues are equal to one in this case. However, these modes are precisely $|\uparrow\uparrow\rangle$ and $|\downarrow\downarrow\rangle$, hence once again $\delta\sigma=-\frac{e^{2}}{2\pi^{2}}\log\frac{\tau_{\phi}}{\tau}$. 

In contrast, when $h_{z}=0$ and $\lambda k_{F}>\Gamma$, \textbf{U} is diagonalized by the total angular momentum basis $|J,M\rangle$, and the only divergent mode corresponds to the singlet ($|0,0\rangle$) state. It follows that the quantum interference correction changes sign yielding WAL: $\delta\sigma=\frac{1}{2}\frac{e^{2}}{2\pi^{2}}\log\frac{\tau_{\phi}}{\tau}$.
This limit of the M2DEG model also corresponds to long-established theoretical results.

In the most general case both $\lambda k_{F}$ and $h_{z}$ may be comparable to the Fermi energy.  \textbf{U} is then not diagonal in either $|m,m^{\prime}\rangle$ or $|J,M\rangle$ representations. In this case, analytical expressions become cumbersome and it is more convenient to display the results graphically (Figs. 2-5). Fig. ~\ref{fig:sigma_vs_so_M2DEG} shows the competing influences of $h_{z}$ and $\lambda k_{F}$; the former favors WL whereas the latter leads to WAL. This trend may be understood by recognizing that most of the conductivity correction stems from \emph{intra-band} transitions; inter-band interference is suppressed due to band-splitting. When $h_{z} \gg \lambda k_{F}$ ($h_{z}<<\lambda k_{F}$), the spinor at $|\alpha,\textbf{k}\rangle$ is nearly parallel (antiparallel) to the spinor at $|\alpha,-\textbf{k}\rangle$, hence as mentioned in the Sec. I the outcome is WL (WAL). This argument rests crucially in the fact that the Rashba SO interaction reverses spinors under space inversion.\cite{chirality} The crossover between WAL and WL occurs when $\lambda k_{F}\simeq 2 h_{z}$, namely when the exchange splitting and SO splitting are nearly identical. 
\begin{figure}[h]
\begin{center}
\scalebox{0.45}{\includegraphics{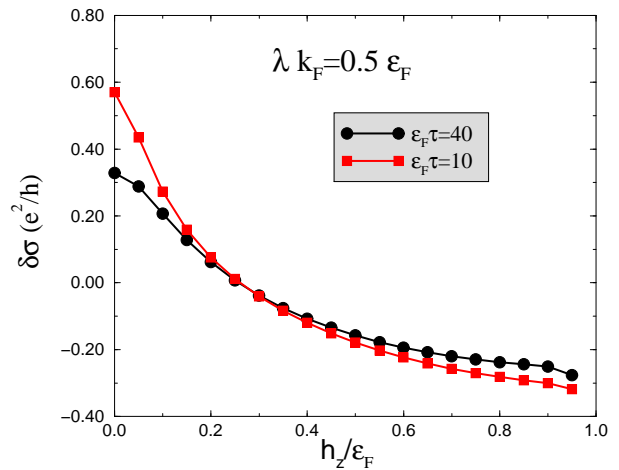}}
\caption{(color online)\textbf{M2DEG}: Exchange field dependence of conductivity correction for $\epsilon_{F}\tau_{\phi}\simeq 840$ and $u^{z}=0$.
Note that the value of the exchange field at which the transition from WAL to WL occurs is independent of the scattering rate. 
This is true because the WAL-to-WL transition in a M2DEG is largely determined by changes in intra-band correlations.}
\label{fig:tau_M2DEG}
\end{center}
\end{figure}

\begin{figure}[h]
\begin{center}
\scalebox{0.45}{\includegraphics{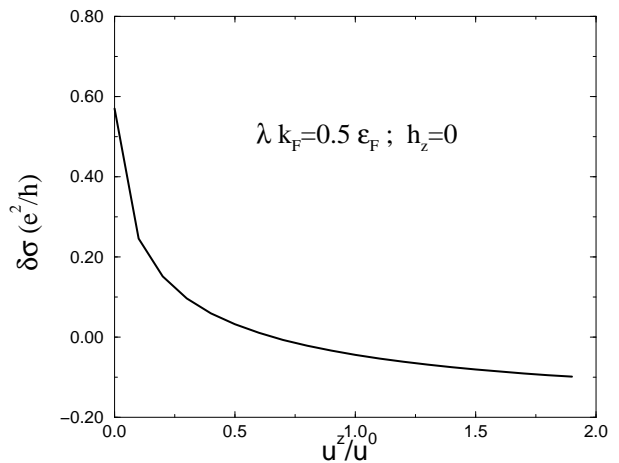}}
\caption{(color online)\textbf{M2DEG}: effect of exchange scatterers in a unmagnetized ($h_z=0$) two-dimensional electron gas with Rashba SO interactions and $\epsilon_{F}\tau_{0}=20$, where $\tau_{0}$
is the scattering time due to non-magnetic impurities.
Exchange impurities dephase the singlet Cooperon responsible for WAL
and leave the triplet Cooperons unchanged. Hence there is a WAL-to-WL transition. The magnitude of WL keeps increasing at larger $u^z$ because the total scattering rate too is increasing. 
}
\label{fig:uz_M2DEG}
\end{center}
\end{figure}

Fig. ~\ref{fig:tau_M2DEG} shows that the value of the exchange field at which the WAL to WL crossover occurs is independent of the scattering rate; this is so because for $h_{z},\lambda k_{F}>\Gamma$, $\delta\sigma$ in a M2DEG is largely governed by intra-band correlations. Fig. ~\ref{fig:uz_M2DEG} studies the effect of magnetic impurities in the  M2DEG. As can be inferred from Eq. (~\ref{eq:U_analytical}), $u^{z}$ dephases opposite spin correlations, yet does not affect equal spin channels.\cite{uz} Consequently, incorporating magnetic impurities may turn WAL into WL.

Table 1 and Figs. ~\ref{fig:sigma_vs_so_M2DEG}- ~\ref{fig:uz_M2DEG} omit the internal magnetic field inherent to ferromagnets. This neglect is justified in thin
film geometries when the magnetization is perpendicular to the 2DEG and the demagnetization field cancels out the internal field. In addition, we have not yet considered the dephasing due to a perpendicular external magnetic field $H_{ext}$. Analyzing the influence of $H_{ext}$ is important because it may uncover non-monotonic MR {\em e.g.} when the magnitudes of the exchange and SO splitting are close to one another. In such scenario a positive (negative) value of $\delta\sigma(H_{ext}=0)$ does {\em not} rule out WL (WAL) for certain range of $H_{ext}$. 
We shall return to these considerations in the next section, where we study the 4-band, kinetic exchange model relevant to (Ga,Mn)As.

\section{Quantum Corrections to Conductivity in $\textbf{(Ga,Mn)As}$}
The goal of this section is to compute the quantum corrections to conductivity for (Ga,Mn)As. Our calculation is motivated in part by the recent controversy\cite{purdue, neumaier, brunocomment,regensbergcomment} concerning the sign of quantum MR in (Ga,Mn)As.  Prior to our work, it was clear that the confusion surrounding this question could be 
reduced by a convincing theoretical answer to the following question:  
is the standard theory of quantum interference in a disordered valence band 
compatible with the observation of negative MR?  Since the SO splitting in the valence band of GaAs ($\Delta_{so}=341$ meV) is larger than the typical exchange splitting ($h_z\simeq 240$ meV at 8\% Mn concentration), the answer would be no, in agreement with Rokhinson {\em et al.},\cite{purdue} if guessed from studies of the M2DEG model described in the previous section. 
However, the detailed calculation for the Kohn-Luttinger kinetic-exchange model performed in this section
 shows that the WL dominates despite the strong SO coupling. The key difference between M2DEGs and (Ga,Mn)As is that the respective quasiparticles have qualitatively different chiralities, which renders WAL significantly more fragile in the latter system than in the former. We elaborate on this point below. 

We adopt the 4-band Kohn-Luttinger Hamiltonian within the spherical approximation using parameters appropriate for GaAs, 
and combine it with a kinetic-exchange mean-field theory model for the ferromagnetic ground state of Mn-doped GaAs:
\begin{equation}
\label{eq:hamiltonian_GaMnAs}
H=\frac{1}{2 m}\left[\left(\gamma_{1}+\frac{5}{2}\gamma_{2}\right)k^{2}-2\gamma_{3}(\textbf{k}\cdot\textbf{J})^{2}\right]+h_{z} J_{z},
\end{equation}
where $\textbf{J}$ is the spin operator projected onto the J=3/2 total angular momentum subspace at the top of the valence band and  \{$\gamma_{1}=6.98, \gamma_{2}=\gamma_{3}=2.5$\} are the Luttinger parameters for the spherical approximation to the valence bands of GaAs. In Eq.(~\ref{eq:hamiltonian_GaMnAs}) $h_{z}=J_{pd} S N_{Mn}$ is the exchange field, $J_{pd}=55 \mbox{meV} \mbox{nm}^{3}$ is the p-d exchange coupling, $S=5/2$ is the spin of Mn ions, $x$ is the Mn fraction, $N_{Mn}=4x/a^{3}$ is the density of Mn ions, and $a=0.565 \mbox{nm}$ is the lattice constant of GaAs.\cite{fsm review} 

Since Eq. (~\ref{eq:hamiltonian_GaMnAs}) cannot be diagonalized analytically (except for $x=0$), Eqs. (~\ref{eq:delta_sigma}), (~\ref{eq:expanded_Cooperon}), (~\ref{eq:matrix equation}) and (~\ref{eq:U_expanded}) must be solved numerically. As in the previous section the central quantity to be computed is \textbf{U}(\textbf{Q}), which in this case is a 16$\times$16 matrix. For $Q=0$, many of its matrix elements vanish due to the rotational symmetry around the exchange field, which renders $U^{m,l}_{m^{\prime},l^{\prime}}(0)\propto \delta_{m+m^{\prime},l+l^{\prime}}$ upon azimuthal integration. As in the previous section, we assume that $U^{m,l}_{m^{\prime},l^{\prime}}(Q)$ is also proportional to $\delta_{m+m^{\prime},l+l^{\prime}}$, which expedites the numerical calculations considerably. In addition, we regard \textbf{Q} as a three-dimensional momentum, i.e., assume that all three dimensions of the (Ga,Mn)As conducting channel are not significantly smaller than $l_{\phi}$.

Let us first consider the case where the exchange-splitting (the density of ordered Mn local moments) is very small. 
In this case the energy spectrum of Eq. (~\ref{eq:hamiltonian_GaMnAs}) consists of nearly degenerate heavy hole bands (HH1 and HH2) and nearly degenerate light hole bands (LH1 and LH2). 
Our numerical studies confirm that there is only one eigenvalue in \textbf{U(0)} that is nearly equal to unity, and that 
it corresponds to the zero total angular momentum (singlet) mode, namely $|0,0\rangle=\frac{1}{2}\left[|\frac{3}{2},-\frac{3}{2}\rangle-|-\frac{3}{2},\frac{3}{2}\rangle+|-\frac{1}{2},\frac{1}{2}\rangle-|\frac{1}{2},-\frac{1}{2}\rangle\right]$.   The conductivity correction from this correlation mode is positive, giving rise to WAL. 
In agreement with Ref. [~\onlinecite{pikus}], we find that the Cooperons correlating HH1 with HH2 and LH1 with LH2 are prominent, 
while the remaining Cooperons are much weaker. This indicates that WAL in (Ga,Mn)As is due to 
\emph{inter-band} interference (HH1-HH2 and LH1-LH2). This scenario is qualitatively different from that of the M2DEG, where WAL is encoded in intra-band correlations.
The source of this crucial difference is that the 4-band Kohn-Luttinger Hamiltonian is invariant under $\textbf{k}\rightarrow -\textbf{k}$, while the M2DEG Hamiltonian is not. 
Accordingly, the spinor $|HH1,\textbf{k}\rangle$ is \emph{parallel} to $|HH1,-\textbf{k}\rangle$ and antiparallel to $|HH2,-\textbf{k}\rangle$;\cite{light holes} as pointed in the previous section, the state of affairs is quite opposite in M2DEGs. Since WAL (WL) follows 
from correlations between spinors pointing in the opposite (same) direction, there are \emph{competing} intra-band (WL) and inter-band (WAL) correlations in (Ga,Mn)As. 
The latter prevail for very low exchange splitting because SO interactions dephase all Cooperon channels except for the singlet.

As the Mn concentration is increased, the HH1-HH2 and LH1-LH2 degeneracies are lifted. Had WAL in GaAs relied mostly on intra-band correlations as in a M2DEG, 
this splitting would not have changed the sign of the conductivity correction until $h_{z}\gtrsim\Delta_{so}$. 
Yet, the WAL contribution in (Ga,Mn)As is sustained by inter-band (HH1-HH2 and LH1-LH2) correlations, which
weaken significantly when the exchange splitting becomes comparable to the \emph{scattering rate} of the quasiparticles. This may be understood by the following simple argument.
When the bands are split, the minimum inter-band momentum sum is $Q_{eff} \sim h_z/v_F$, where $v_F$ is the Fermi velocity .  
Inter-band interference becomes negligible when $D Q_{eff}^2 \sim 1/\tau$, {\em i.e.} when $h_z \sim 1/\tau$.    
Meanwhile, the intra-band correlations (which favor WL) have zero center-of-mass momentum regardless of the splitting.
Furthermore, since oppositely directed momenta already have parallel spinors in the absence of a field, these 
intra-band correlations are insensitve to exchange splitting.  This behavior also contrasts with the M2DEG behavior
in which intra-band WL correlations are enhanced by $h_z$.
As a consequence the crossover from inter-band dominated to intra-band dominated interference 
occurs already when $h_{z}\lesssim 1/\tau$ in (Ga,Mn)As, a condition which is 
qualitatively less stringent than the $h_{z}\gtrsim\Delta_{SO}$ condition that applies 
in M2DEGs and in other systems with similar chirality.

\begin{figure}[h]
\begin{center}
\scalebox{0.45}{\includegraphics{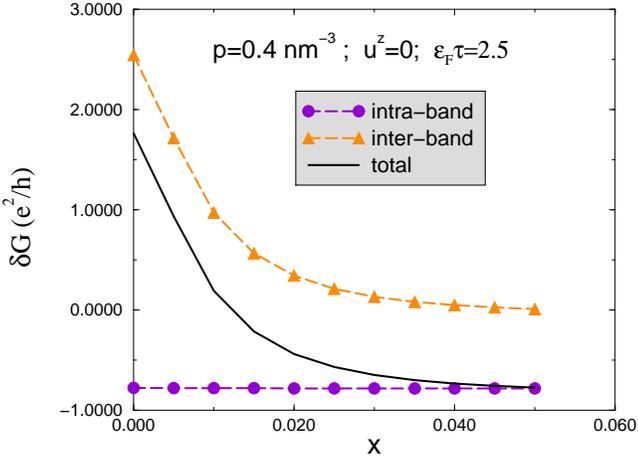}}
\caption{(color online)\textbf{(Ga,Mn)As}: Inter-band vs. intra-band conductance correction {\em vs.} Mn fraction $x$ for $l_{\phi}\simeq 100 nm$ and $u^{z}=0$ in absence of an external field. $p$ stands for the hole-density.
In this figure we have set inter-band matrix elements of the velocity operator to zero in order
to make the distinction between inter-band and intra-band contributions to quantum interference clearer. 
The inter-band ($\alpha\neq\alpha^{\prime}$) correlations responsible for WAL in GaAs decay rapidly as the exchange field due to Mn ions lifts band degeneracies. 
In contrast, intra-band ($\alpha=\alpha^{\prime}$) correlations are relatively indifferent to exchange splitting, and are responsible for the eventual crossover to WL. 
For $p=0.4 nm^{-3}$ and $\epsilon_{F}\tau=2.5$, $h_z=1/\tau$ is satisfied at $x=0.05$ -- note that inter-band correlations are nearly vanished at that point. Assuming a thin film geometry with square cross section, the conductance in this figure is defined as $G=\sigma t$, where $\sigma$ is the calculated conductivity and $t$.
is the thickness of the film (100 nm).}

\label{fig:intra_vs_inter}
\end{center}
\end{figure}
\begin{figure}[h]
\begin{center}
\scalebox{0.45}{\includegraphics{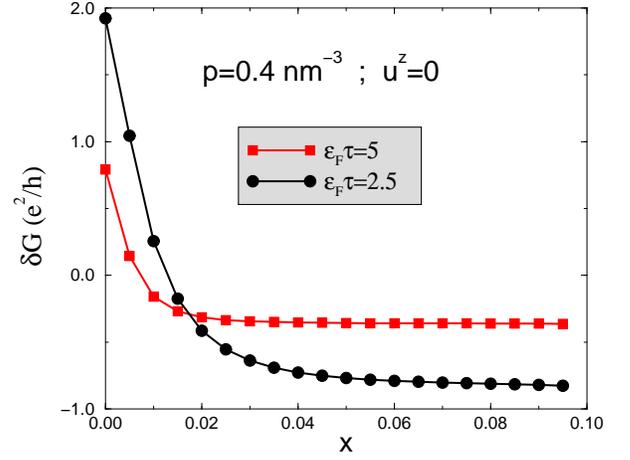}}
\caption{(color online)\textbf{(Ga,Mn)As}: Total conductance correction for two different values of 
disorder. Since the inter-band contribution which favors positive conductance 
contributions vanishes when $h_z \sim 1/\tau$, the value of the Mn concentration at which the WAL-to-WL crossover occurs depends on the scattering rate.
This behavior is different from that of the M2DEG model.
}
\label{fig:sigma_vs_u0}
\end{center}
\end{figure}
\begin{figure}[h]
\begin{center}
\scalebox{0.45}{\includegraphics{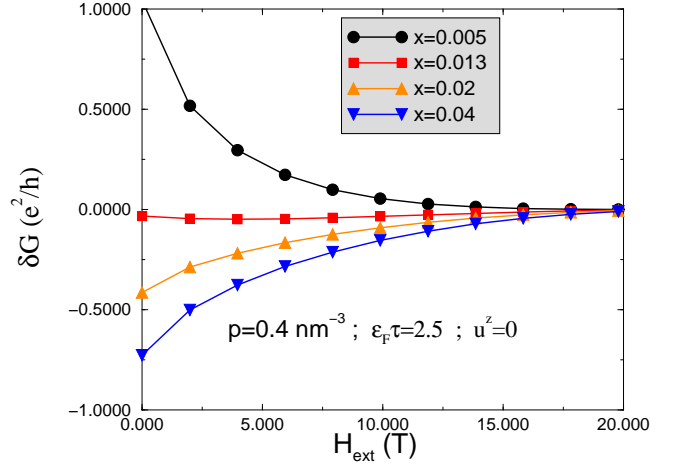}}
\caption{(color online)\textbf{(Ga,Mn)As}: Magnetoresistance for $u^{z}=0$ for several different Mn concentrations.
The short mean-free-paths of (Ga,Mn)As imply that the magnetoresistive signal persists up to relatively large fields. 
The conductance is defined as $G=\sigma t$, where $\sigma$ is the calculated conductivity and $t$ is the thickness of a 
square film (100 nm). At very small $x$, we find the positive MR characteristic of WAL; the sign changes as a function of Mn concentration.}
\label{fig:MR}
\end{center}
\end{figure}
\begin{figure}[h]
\begin{center}
\scalebox{0.45}{\includegraphics{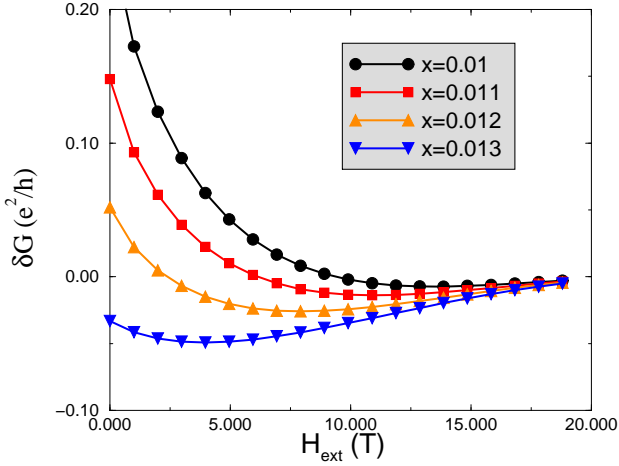}}
\caption{(color online)\textbf{(Ga,Mn)As}: Magnetoresistance for $u^{z}=0, p=0.4 nm^{-3}, \epsilon_F \tau=2.5$  at several different Mn concentrations near the crossover from negative to positive 
MR. }
\label{fig:MR2}
\end{center}
\end{figure}

\begin{figure}[h]
\begin{center}
\scalebox{0.45}{\includegraphics{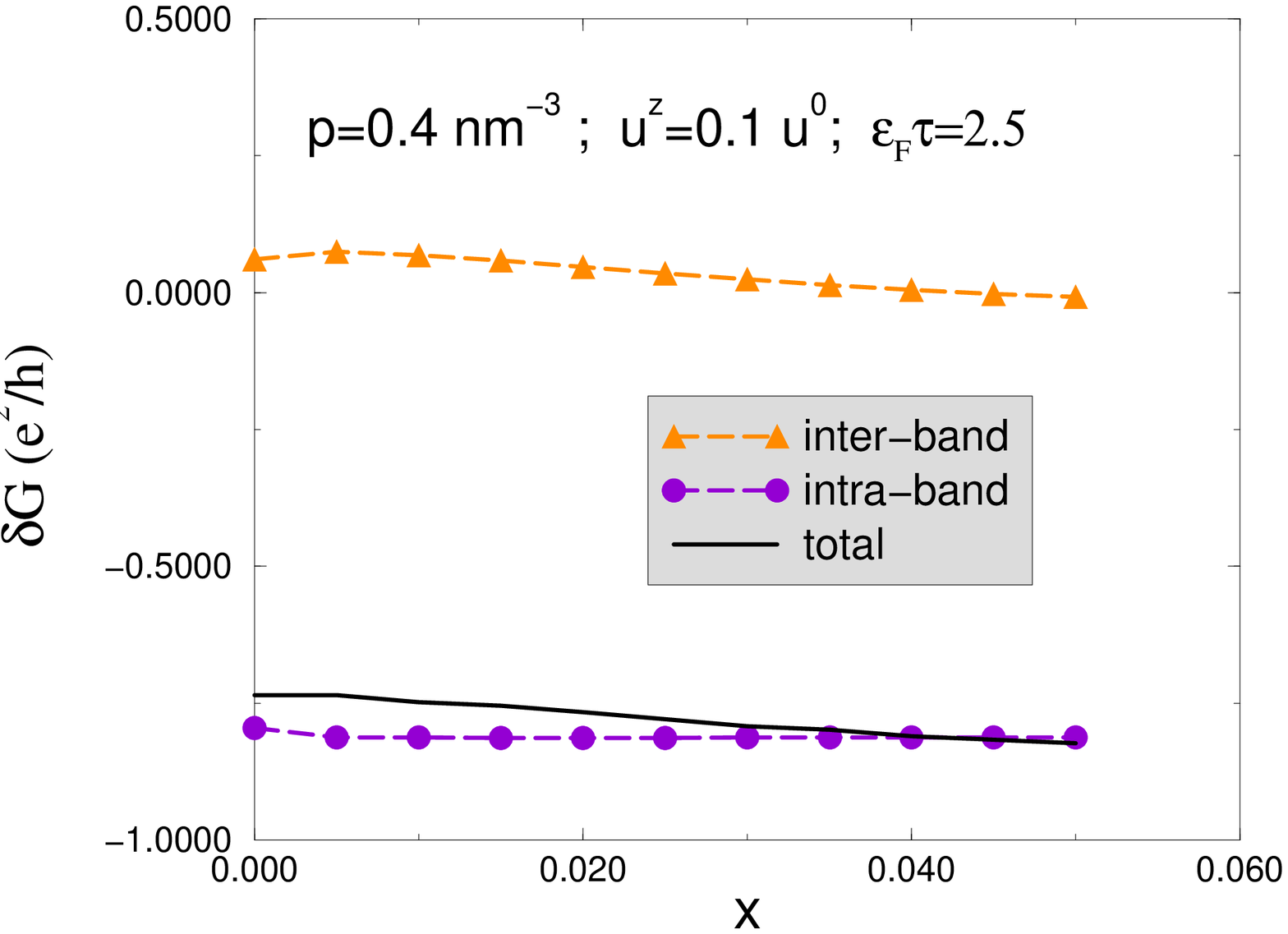}}
\caption{(color online)\textbf{(Ga,Mn)As}: Exchange impurities reverse the sign of magnetoresistance in GaAs by decreasing the WAL correlations while leaving the WL correlations unaffected. $\tau_{0}$ is the scattering time off non-magnetic impurities.}
\label{fig:uz_GaMnAs}
\end{center}
\end{figure}

Figs. ~\ref{fig:intra_vs_inter} and ~\ref{fig:sigma_vs_u0} illustrate the preceding observations. Fig. ~\ref{fig:intra_vs_inter} demonstrates the competition between inter-band (WAL) and intra-band (WL) correlations and the crossover that occurs as band degeneracies are lifted by the exchange field. On the other hand, Fig. ~\ref{fig:sigma_vs_u0} shows that the crossover 
between positive and negative conductance corrections occurs at smaller Mn concentration for cleaner samples in which the 
inter-band contribution is destroyed by weaker exchange fields.  The saturation of the magnitude of $\delta\sigma$ at larger $x$ reflects a nearly complete decay of inter-band correlations and the
weak dependence of intra-band correlations on $h_z$. 

It follows that negative MR is not only possible in (Ga,Mn)As, but expected through most of the 
metallic regime to which this theory is intended to apply.  To address the question of whether positive MR can occur 
in any parameter and field range, even when the total conductivity change is negative, we investigate the influence of an external magnetic field on the quantum correction to conductivity.
This task is most simply (yet not most accurately) accomplished by smoothly replacing the coherence length $l_{\phi}$ by the magnetic length $l_{H}=\left(\hbar/(e H_{ext})\right)^{1/2}$
through the substitution of $1/l_\phi$ by $\left(l_\phi^{-2}+ l_H^{-2}\right)^{1/2}$.
The outcome is Fig. ~\ref{fig:MR}, which showcases the transition from positive to negative MR as a function of Mn concentration. 
The effect of the external field becomes significant when $l_{H}$ is smaller than an \emph{effective} coherence length $l_{\phi}^{eff}$ which will be smaller than $l_\phi$ due to the 
SO coupling dephasing and effects of dephasing due to the exchange splitting. 
For no exchange splitting, $x=0$, $l_{\phi}^{eff}=l_{\phi}\simeq 100$ nm 
and the suppression of WAL becomes visible as $H_{ext}\gtrsim 0.1 T$. For $x>0$, the competition between the exchange field and the SO interaction removes the singularity from the singlet channel. The removal of this singularity is qualitatively akin to reducing the decoherence length with respect to the $x=0$ value, i.e. $l_{\phi}^{eff}<l_{\phi}$. Accordingly, a larger $H_{ext}$ is required to make a dent in the quantum correction to conductivity, thereby yielding shallower MR curves.\cite{comparison} At any rate, the quantum correction to conductivity and its subsequent suppression under an external field ought to be observable so long as $l_{\phi}^{eff}>l$.
We note that the substitution of $1/l_\phi$ by $(l_\phi^{-2}+l_H^{-2})^{1/2}$ is not quantitatively accurate for $l_{H}>l_{\phi}$ ($0 < H_{ext}\lesssim 0.06 T$); yet as we show below this does not erradicate the possibility of WAL from our theory.  In viewing all these figures it should be kept in mind that an
impurity band is expected\cite{tomasib} to form in (Ga,Mn)As for $x \lesssim 0.01$, close to but possibly at larger 
values of $x$ than the MIT.  Our model results in this parameter range should be viewed with caution.
In Fig. ~\ref{fig:MR2} we take a closer look at the doping region where $\delta\sigma(H_{ext}=0)$ changes sign (which is this regime) 
and we find a non-monotonic magnetoresistance. In spite of some qualitative similarities, these results are quantitatively distinct from Neumaier {\em et al.}'s measurements: our WAL ``bumps'' are less pronounced and they disappear as doping increases.  More to the point, the samples studied by Neumaier {\em et al.} have larger values of $x$.

Finally, Fig. ~\ref{fig:uz_GaMnAs} highlights the effect of spin-dependent scatterering, which is inevitable due to the random distribution Mn moments.
As in the case for the M2DEG, magnetic scatterers dephase WAL correlations mainly, further reducing the likelihood of the appearance of positive MR.

\section{Anisotropic Weak Localization}
Most experimental queries of quantum interference in (Ga,Mn)As involve external magnetic fields that are misaligned with the easy axis. Consequently, experimenters attempt to identify and subtract a background made of AMR. However, the anisotropic {\em quantum interference} that accompanies the normal AMR receives little mention; in this section we study its possible implications.

While there are a variety of crystalline and non-crystalline sources for AMR, in the spherical 4-band model that we study in this paper only the non-crystalline term given by the relative angle between the magnetization and current is non-zero. The  non-crystalline AMR in (Ga,Mn)As has been shown to stem from scattering from Mn impurities whose magnetic and Coulomb potentials are treated coherently when evaluating the anisotropic lifetimes.\cite{amr} In order to allow for scattering centers that have correlated spin-dependent and spin-independent parts, the formalism introduced in Section III must be modified slightly; we relegate the details to Appendix B and instead present the results directly. 

\begin{figure}[h]
\begin{center}
\scalebox{0.45}{\includegraphics{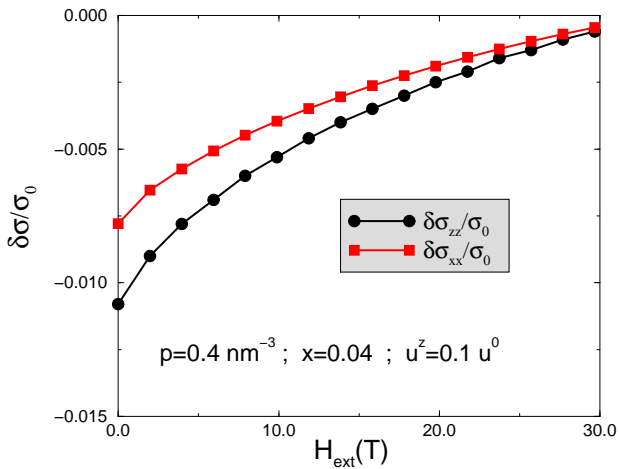}}
\caption{(color online) Dependence of magnetoresistance on the relative orientation between the electric current and the magnetic easy axis ($\hat{z}$) in (Ga,Mn)As. $\sigma_0=(\sigma_{xx}+\sigma_{zz})/2$ is the average value of the Boltzmann conductivity.}
\label{fig:q_amr}
\end{center}
\end{figure}

Fig. ~\ref{fig:q_amr} reveals a sizeable ($\gtrsim$ 20\%) anisotropy in the quantum corrections to conductivity which is similar percentage wise to the anisotropy in the Boltzmann conductivity. We find that the Boltzmann conductivity is larger when magnetization is along the current due to the anisotropy of the scattering lifetime; likewise, the WL correction is enhanced when current and magnetization are parallel. This leads to a smaller absolute change in the total conductivity when magnetization is rotated from along to perpendicular to current in the presence of the WL but in the same time the conductivity itself is also suppressed due to the WL. The resulting relative AMR ratio is therefore nearly independent on whether the WL corrections are included or not.
These results agree qualitatively with analytical considerations on simpler models.\cite{awl} 

In the experiment by Neumaier {\em et al.}\cite{neumaier} the relative AMR ratio seems to depend on temperature and, more importantly, the shape of the low-field MR curve develops additional features when lowering the temperature. The anisotropic WL effects we calculated cannot account for these observations.

\section{Conclusions}
Motivated by the recent experimental studies of the influence of quantum interference on the transport properties of (Ga,Mn)As, we have presented a theory of quantum 
interference corrections to conductivity in disordered ferromagnets with intrinsic SO interaction. We have focused on two simple models, a toy two-dimensional electron gas ferromagnet with Rashba SO interaction and a 4-band Kohn-Luttinger kinetic-exchange model, the latter of which provides an approximate representation of (Ga,Mn)As.  Our results for (Ga,Mn)As show that in spite of the very strong SO coupling of 
holes in the valence band, a negative MR is not only possible but also overwhelmingly likely. Starting from vanishing exchange fields, {\em i.e.} vanishing Mn concentration, we find a transition from WAL to WL as the Mn local moment concentration is increased.  This transition occurs at relatively low doping concentrations, close to the 
values at which the disordered valence band model of (Ga,Mn)As starts to become valid.  The transition occurs at 
weak exchange fields because the chirality of the (Ga,Mn)As quasiparticles encodes WAL in inter-band correlations, which decay substantially as the exchange splitting becomes comparable to the scattering rate off impurities.

Although our theory does allow for the possibility of a non-monotonic MR at small Mn fractions, near the theory's validlity limits,
it does not explain the character of the positive MR observed in the experiment of Neumaier {\em et al.}\cite{neumaier} and therefore does not confirm the WAL interpretation of the data.  
It is desirable to generalize our calculation to more accurate models that account for departures from the 
spherical approximation and include the two split-off valence bands, strain and electron-electron interactions.
Nevertheless we believe that the present calculation has uncovered the essence of WL/WAL physics in (Ga,Mn)As and explained 
essential differences between this system and the M2DEG model studied previously.
Since the 4-band Kohn-Luttinger kinetic-exchange model studied in this paper represents the high SO coupling limit of the more accurate 6-band Kohn-Luttinger Hamiltonian, the dominance of the WL we obtained here is unlikely to be suppressed by the inclusion of higher bands. 

\section*{Acknowledgment}
We thank K. Vyborny, V. Fal'ko, B. Gallagher and J. Wunderlich for fruitful conversations.
This work was supported by
ONR under Grant No. onr-n000140610122, by
NSF under Grant No. DMR-0547875, by SWAN-NRI, by the Welch Foundation, 
by EU No. IST-015728, FP7  No. 214499 (NAMASTE), Praemium Academiae, Czech Republic No. AV0Z1-
010-914, No. KAN400100625, No. LC510, and No.
FON/06/E002. Jairo Sinova is a Cottrell Scholar of the
Research Corporation.

\appendix
\begin{widetext}
\section{Derivation of Table 1}
Starting from Eq. (~\ref{eq:U_analytical}) we derive the results shown in Table 1.

\subsection{$h_{z}=0$,$\lambda=0$}

In this case $U^{\uparrow\uparrow}_{\uparrow\uparrow}=U^{\downarrow\downarrow}_{\downarrow\downarrow}=1, U^{\uparrow\downarrow}_{\downarrow\uparrow}=U^{\downarrow\uparrow}_{\uparrow\downarrow}=0, U^{\uparrow\uparrow}_{\downarrow\downarrow}=U^{\downarrow\downarrow}_{\uparrow\uparrow}=(u^{0}-u^{z}/4)/(u^{0}+u^{z}/4)$. Then the Cooperon is diagonal in the $J^{z}$ representation:n
\begin{equation}
\textbf{C}=(\textbf{1-U})^{-1}\textbf{C}^{(0)}=\left(\begin{array}{cccc} C^{\uparrow\uparrow}_{\uparrow\uparrow} & 0 & 0 & 0\\
                                              0 & C^{\uparrow\uparrow}_{\downarrow\downarrow} & 0 & 0\\
                                              0 & 0 & C^{\downarrow\downarrow}_{\uparrow\uparrow} & 0\\
                                              0 & 0 & 0 & C^{\downarrow\downarrow}_{\downarrow\downarrow}
\end{array}\right)
\end{equation}
where
\begin{eqnarray}
&&C^{\uparrow\uparrow}_{\uparrow\uparrow}=C^{\downarrow\downarrow}_{\downarrow\downarrow}=\frac{1}{2\pi N_{2D}\tau}\frac{1}{D Q^{2}\tau}\nonumber\\
&&C^{\uparrow\uparrow}_{\downarrow\downarrow}=C^{\downarrow\downarrow}_{\uparrow\uparrow}=\frac{(u^{0})^{2}-(u^{z})^{2}/16}{u^{z}/2+(u^{0}-u^{z}/4)DQ^{2}\tau}
\end{eqnarray}
Recalling Eq. (~\ref{eq:delta_sigma}), the quantum correction to conductivity reads 
\begin{equation}
\label{eq:delta_sigma_0}
\delta\sigma=\frac{e^{2}}{2\pi}\int_{\textbf{k}}v_{\alpha,\alpha}^{x}(\textbf{k})v_{\alpha^{\prime},\alpha^{\prime}}^{x}(-\textbf{k})G_{\alpha}^{R}(\textbf{k})G_{\alpha^{\prime}}^{R}(\textbf{k})G_{\alpha}^{A}(\textbf{k})G_{\alpha^{\prime}}^{A}(\textbf{k})\int_{\textbf{Q}}C^{\alpha\alpha^{\prime}}_{\alpha^{\prime}\alpha}(\textbf{k},\textbf{Q})
\end{equation}
\end{widetext}
where we have used the fact that the velocity operators are diagonal in spin space. It is clear from Eq. (~\ref{eq:delta_sigma_0}) that $C^{\uparrow\uparrow}_{\downarrow\downarrow}$ and $C^{\uparrow\uparrow}_{\downarrow\downarrow}$ do not contribute to $\delta\sigma$; the physical significance of this has been explained in Section III. 
\begin{eqnarray}
\delta\sigma&=&-\frac{e^{2}}{2\pi}\int_{\textbf{k}}v_{x}^{2}(G_{R})^{2}(G_{A})^{2}\int_{\textbf{Q}}\left(C^{\uparrow\uparrow}_{\uparrow\uparrow}+C^{\downarrow\downarrow}_{\downarrow\downarrow}\right)\nonumber\\
&=& -\frac{e^{2}}{2\pi^{2}}\log\frac{\tau_{\phi}}{\tau}
\end{eqnarray}
where we have used $Q_{\text{min}}=(D\tau_{\phi})^{-1/2}$ and $Q_{\text{max}}=(D\tau)^{-1/2}$. Remarks: (i) the exchange impurities have no effect in this case, because they do not dephase Cooperons with maximal projection of angular momentum; (ii) for $0<h_{z}<\Gamma$ we'd get the same answer to order $0(1/(h_{z}\tau))$.

\subsection{$\Gamma<h_{z}<\epsilon_{F}$, $\lambda=0$}

Here $U^{\uparrow\uparrow}_{\uparrow\uparrow}=U^{\downarrow\downarrow}_{\downarrow\downarrow}=1, U^{\uparrow\downarrow}_{\downarrow\uparrow}=U^{\downarrow\uparrow}_{\uparrow\downarrow}=0, U^{\uparrow\uparrow}_{\downarrow\downarrow}=-U^{\downarrow\downarrow}_{\uparrow\uparrow}=-(u^{0}-u^{z}/4) i\pi N_{2D}/h_{z}$. $U^{\uparrow\uparrow}_{\downarrow\downarrow}$  and $U^{\downarrow\downarrow}_{\uparrow\uparrow}$ are different from the previous case; however, since $C^{\uparrow\uparrow}_{\downarrow\downarrow}$  and $C^{\downarrow\downarrow}_{\uparrow\uparrow}$ do not contribute to the conductivity correction, once again we arrive at
\begin{equation}
\delta\sigma= -\frac{e^{2}}{2\pi^{2}}\log\frac{\tau_{\phi}}{\tau}
\end{equation}
Therefore, an exchange field {\it per se} does not decrease the WL correction to the conductivity. However, we have neglected the orbital effect due to the magnetization of the ferromagnet; if strong enough, this may entirely suppress quantum interference.

\subsection{$h_{z}=0$, $0<\lambda k_{F}<1/(\tau\tau_{\phi})^{1/2}<\Gamma$}

In this case $U^{\uparrow\uparrow}_{\uparrow\uparrow}=U^{\downarrow\downarrow}_{\downarrow\downarrow}=1, U^{\uparrow\downarrow}_{\downarrow\uparrow}=U^{\downarrow\uparrow}_{\uparrow\downarrow}=0, U^{\uparrow\uparrow}_{\downarrow\downarrow}=U^{\downarrow\downarrow}_{\uparrow\uparrow}=(u^{0}-u^{z}/4)/(u^{0}+u^{z}/4)$. Hence we have the same $U$-matrix as in the first case; however, the eigenstates are different now. Then,
\begin{widetext}
\begin{eqnarray}
\delta\sigma &\simeq & \frac{e^{2}}{2\pi}\int_{\textbf{k}}v_{\alpha,\beta}^{x}(\textbf{k}) v_{\beta^{\prime},\alpha^{\prime}}^{x}(-\textbf{k}) G_{\alpha}^{R}(\textbf{k})G_{\alpha^\prime}^{R}(\textbf{k})G_{\alpha}^{A}(\textbf{k})G_{\alpha^\prime}^{A}(\textbf{k})\int_{\textbf{Q}}C^{\beta\beta^{\prime}}_{\alpha^{\prime}\alpha}(\textbf{k},\textbf{Q})\nonumber\\
&\simeq & \frac{e^{2}}{2\pi}\int_{\textbf{k}} v_{\alpha\alpha}^{x}(\textbf{k})v_{\alpha^{\prime}\alpha^{\prime}}^{x}(-\textbf{k}) (G^{R})^{2} (G^{A})^{2}\int_{\textbf{Q}} C^{\alpha\alpha^{\prime}}_{\alpha^{\prime}\alpha}(\textbf{k},\textbf{Q})
\end{eqnarray}
where we have used $v_{\alpha\beta}\propto\delta_{\alpha\beta}$, which is a good approximation provided that $\lambda k_{F}<<\epsilon_{F}$. Then
\begin{eqnarray}
\delta\sigma &\simeq& -\frac{e^{2}}{2\pi}\int_{\textbf{k}}v_{x}^{2}(G^{R})^{2}(G^{A})^{2}\int_{\textbf{Q}}\left(C^{+,+}_{+,+}+C^{+,-}_{-,+}+C^{-,+}_{+,-}+C^{-,-}_{-,-}\right)\nonumber\\
&-& \frac{e^{2}}{2\pi}\int_{\textbf{k}}v_{x}^{2}(G_{R})^{2}(G_{A})^{2}\int_{\textbf{Q}}\left(C^{\uparrow\uparrow}_{\uparrow\uparrow}+C^{\downarrow\downarrow}_{\downarrow\downarrow}\right)\nonumber\\
&=& -\frac{e^{2}}{2\pi^{2}}\log\frac{\tau_{\phi}}{\tau}
\end{eqnarray}
\end{widetext}
where we have used $C^{+,+}_{+,+}=C^{-,-}_{-,-}=\frac{1}{2}(C^{\uparrow\uparrow}_{\uparrow\uparrow}-C^{\downarrow\downarrow}_{\downarrow\downarrow})$ and $C^{+,-}_{-,+}=C^{-,+}_{+,-}=\frac{1}{2}(C^{\uparrow\uparrow}_{\uparrow\uparrow}+C^{\downarrow\downarrow}_{\downarrow\downarrow})$ (these relations can be derived straightforwardly by a basis transformation). We thus conclude that when the SO interaction weaker than the spin-orbit dephasing rate there is no effect of SO in the quantum correction to conductivity. However, if $\lambda k_F\gtrsim1/(\tau\tau_{\phi})^{1/2}$ it can be shown from Eq. (~\ref{eq:U_analytical}) that the dephasing of the triplet Cooperon is no longer negligible, which leads to a smaller magnitude of WL. While the actual analytical expressions are cumbersome in this case, the influence of spin-orbit dephasing may be roughly captured by replacing $l_{\phi}$ with the spin-orbit length $l_{so}=1/(m\lambda)$ in the lower bound of the Q-integral.

\subsection{$h_{z}=0$ and $\Gamma<\lambda k_{F}<\epsilon_{F}$}

Since the band splitting is larger than the scattering rate, we shall neglect the contributions from inter-band transitions. In this case
$U^{\uparrow\uparrow}_{\uparrow\uparrow}=U^{\downarrow\downarrow}_{\downarrow\downarrow}=1/2, U^{\uparrow\downarrow}_{\downarrow\uparrow}=U^{\downarrow\uparrow}_{\uparrow\downarrow}=-1/2(u^{0}-u^{z}/4)/(u^{0}+u^{z}/4), U^{\uparrow\uparrow}_{\downarrow\downarrow}=U^{\downarrow\downarrow}_{\uparrow\uparrow}=1/2(u^{0}-u^{z}/4)/(u^{0}+u^{z}/4)$. Clearly, the Cooperon is no longer diagonal in the $\{|m,m^{\prime}\rangle\}$ basis; instead it is diagonal in the $\{|J,M\rangle\}$ basis, where $J=0,1$, $M=0,\pm 1$. We arrive at
\begin{eqnarray}
C_{1,1}&=&C_{1,-1}=\frac{1}{2\pi N_{2D}\tau}\frac{1}{\frac{1}{2}(1+D Q^{2}\tau)}\nonumber\\
C_{0,0}&=&\frac{1}{2\pi N_{2D}\tau}\frac{1}{1-\frac{u^{0}-u^{z}/4}{u^{0}+u^{z}/4}(1-D Q^{2}\tau)}
\end{eqnarray}
where we have ignored $C_{1,0}$ because it will not contribute to $\delta\sigma$. Note that the triplet channel ($C_{1,M}$) is dephased, while the singlet channel ($C_{0,0}$) is singular provided that there are no exchange impurities. For simplicity,  let us take $u^{z}\simeq 0$. Then
\begin{equation}
C_{0,0}\simeq\frac{1}{2\pi N_{2D}\tau}\frac{1}{D Q^{2}\tau}
\end{equation}
Consequently, the correction to conductivity reads
\begin{equation}
\delta\sigma\simeq \frac{e^{2}}{2\pi}\int_{\textbf{k}}v_{\alpha\alpha}^{x}(\textbf{k})v_{\alpha\alpha}^{x}(-\textbf{k})(G_{\alpha}^{R})^{2}(G_{\alpha}^{A})^{2}\int_{\textbf{Q}}C^{\alpha\alpha}_{\alpha\alpha}
\end{equation}
With further basis transformations, one can show that $C^{+,+}_{+,+}=C^{-,-}_{-,-}=\frac{1}{2}(C_{1,1}-C_{0,0})\simeq-\frac{C_{0,0}}{2}$, with which we reach
\begin{equation}
\delta\sigma=\frac{e^{2}}{4\pi^{2}}\log\frac{\tau_{\phi}}{\tau}
\end{equation}
This is nothing but WAL, with a magnitude that is half of the traditional WL.

\section{Anisotropic Weak Localization}
In Section III we have developed a procedure to evaluate the Cooperon in arbitrary, SO coupled ferromagnets with incoherent magnetic and Coulomb scatterers. In this appendix we generalize such procedure so as to allow for coherent charge- and spin-scatterers.

On one hand, the new Bethe-Salpeter equation for the Cooperon reads 
\begin{widetext}
\begin{equation}
C^{\beta,\beta^{\prime}}_{\alpha^{\prime},\alpha}=u^{a,b} J^{a}_{\beta,\beta^{\prime}} J^{b}_{\alpha^{\prime},\alpha}+\int_{{\bf k}''} u^{a,b} J^{a}_{\beta,\beta''} J^{b}_{\alpha',\alpha''} G^{A}_{\beta''}G^{R}_{\alpha''} C^{\beta'',\beta'}_{\alpha'',\alpha}
\end{equation}
where we have omitted the momenta labels and  $u^{0,0}=u^{0}$, $u^{z,z}=u^z$,$u^{0,z}=u^{z,0}=\sqrt{u^0 u^z}$ are the only non-zero elements of $u^{a,b}$. Carrying out the same transformations as in Section III, we arrive at
\begin{equation}
C^{m,n}_{m',n'}({\bf Q})=(u^0+u^z m m'+\sqrt{u^0 u^z}(m+m'))\delta_{m,n}\delta_{m',n'}+\sum_{l,l'}U^{m,l}_{m',l'}C^{l,n}_{l',n'}
\end{equation}
where
\begin{equation}
U^{m,l}_{m',l'}=\left(u^0+u^z m m'+\sqrt{u^0 u^z}(m+m')\right)\int_{{\bf k}}\langle m|\beta,{\bf k}\rangle G^{A}_{\beta}\langle\beta,{\bf k}|l\rangle\langle m'|\alpha,-{\bf k+Q}\rangle G^R_{\alpha}\langle\alpha'',{\bf -k+Q}|l'\rangle.
\end{equation}
On the other hand, the expression for the transport lifetime is now
\begin{equation}
\frac{1}{\tau_{{\bf k},\alpha}}=2\pi\int_{{\bf k'}} u^{a,b}({\bf k-k'})\sum_{\alpha'} J^{a}_{\alpha,\alpha'} J^{b}_{\alpha',\alpha}\delta(E_{{\bf k},\alpha}-E_{{\bf k'},\alpha'})(1-\hat {\bf k}\cdot\hat {\bf k'})
\end{equation}
where the factor $(1-\hat {\bf k}\cdot\hat {\bf k'})$ stems from the renormalization of the velocity vertex.\cite{awl} 
\end{widetext}


\begin{thebibliography}{50}
\bibitem{footnote}  The AMR contribution to the conductance $G$ in Ref.\onlinecite{neumaier} is $\sim 0.006 e^2/h$ per wire.
The authors assume that the AMR contribution scales with the high field conductance and therefore that 
it is reduced by about 10\% between 65 mK and 20 mK.  This $\sim 0.0006 e^2/h$ reduction in AMR 
$G$ is a significant ($\simeq$ 30 \%) fraction of the conductance peak interpreted as due to WAL.
\bibitem{dimensions} Another possible culprit for the early saturation is dimensionality crossover induced by an external magnetic field. However, one difficulty of this explanation is that even at zero fields the samples are not clearly in the 1D limit.
\bibitem{other_bases}  The $J^{z}$ basis representation choice is motivated by convenience - it diagonalizes the impurity potential we consider.
Choosing $J^{x}$ or $J^{y}$ representations would eventually lead to the same answer.
\bibitem{wrong_hope} However, it will turn out that there are qualitative differences between the 
M2DEG model and more realistic models when it comes to evaluating quantum corrections to conductivity in magnetic semiconductors (see Section IV).
\bibitem{chirality} We have verified, for example, that when the chirality of the Rashba SO is doubled 
(i.e. when spinors rotate twice as fast along the Fermi circle), the singular mode would correspond to $|1,0\rangle$ (triplet) rather than $|0,0\rangle$ (singlet).
 Consequently, WAL is replaced by WL.
\bibitem{comparison} The characteristic $H_{ext}$ at which quantum interference decays in Fig. (~\ref{fig:MR}) is much larger than that of Ref. [~\onlinecite{purdue}], where the WL suppression is evident at as low a field as 0.03 T. Our results are more in tune with Refs. [~\onlinecite{neumaier}] and [~\onlinecite{matsukura}], where the decay of quantum corrections is observed at the level of several Tesla.
\end{thebibliography}
\end{document}